\documentclass{aa}
\usepackage{graphicx}
\usepackage{txfonts}
\usepackage{natbib}
\bibpunct{(}{)}{;}{a}{}{,}

\begin{document}

\title{A Sino-German $\lambda$6~cm polarization survey of the Galactic plane}
\subtitle{IV. The region from $60\degr$ to $129\degr$ longitude}

\author{
L. Xiao\inst{1}
        \and J. L. Han\inst{1}
        \and W. Reich\inst{2}
        \and X. H. Sun\inst{1}
        \and R. Wielebinski\inst{2}
        \and P. Reich\inst{2}
        \and H. Shi\inst{1}
        \and O. Lochner\inst{2}
}

\offprints{J.L. Han}

\institute{National Astronomical Observatories, Chinese Academy of
  Sciences, Jia-20, Datun Road, Chaoyang District, Beijing 100012,
  China\\ \email{hjl@bao.ac.cn} 
\and Max-Planck-Institut f\"{u}r
  Radioastronomie, Auf dem H\"ugel 69, 53121 Bonn, Germany\\
}

\date{Received / Accepted}

\abstract
{Linear polarization of diffuse Galactic emission is a signature of
  magnetic fields in the interstellar medium of our Galaxy.
  Observations at high frequencies are less affected by Faraday
  depolarization than those at lower frequencies and are able to
  detect polarized emission from more distant Galactic regions.}
{We attempt to perform a sensitive survey of the polarized emission
  from the Galactic disk at $\lambda$6~cm wavelength.}
{We made polarization observations of the Galactic plane using the
  Urumqi 25-m telescope at $\lambda$6~cm covering the area of
  $60\degr\leq l \leq129\degr$ and $|b|\leq5\degr$.  Missing
  large-scale structures in polarization were restored by
  extrapolation of the WMAP polarization data.}
{We present the $\lambda$6~cm total intensity and linear polarization maps
  of the surveyed region. We identify two new extended \ion{H}{II}
  regions G98.3$-$1.6 and G119.6+0.4 in this region. Numerous
  polarized patches and depolarization structures are visible in the
  polarization maps. Depolarization along the periphery of a few
  \ion{H}{II} complexes was detected and can be explained by a Faraday
  screen model. We discuss some prominent depolarization \ion{H}{II}
  regions, which have regular magnetic fields of several
  $\mu$G. Structure functions of $U$, $Q$, and $PI$ images of the
  entire $\lambda$6~cm survey region of $10\degr\leq l \leq 230\degr$
  exhibit much larger fluctuation power towards the inner Galaxy,
  which suggests a higher turbulence in the arm regions of the
  inner Galaxy.}
{The Sino-German $\lambda$6~cm survey reveals new properties of the
  diffuse magnetized interstellar medium. The survey is also very
  useful for studying individual objects such as \ion{H}{II} regions,
  which act as Faraday screens and have high rotation measures and therefore
  strong regular magnetic fields inside the regions.}


\maketitle

\section{Introduction}

A polarization survey of diffuse Galactic emission provides a direct
image of the transverse distribution of the magnetic field, if Faraday
rotation is negligible. In all polarization surveys, depolarization
occurs by vector canceling of polarized emission in a beam or from
different layers along the line-of-sight \citep{burn66,sbs98}.  
Observations at different frequencies show the polarized emission at
different Faraday depths \citep[e.g. ][]{wdj93,hkd03a,hkd03b,ulg03}.
Polarization surveys reveal numerous small-scale polarized structures
unrelated to total intensity emission. Patches having polarization angle
deviations compared to their surroundings are interpreted as Faraday
screens \citep{gld1998}, which do not emit any synchrotron emission
but rotate the polarization angles of background emission. These
Faraday screens could be either \ion{H}{II} regions \citep[e.g. ][]{shr07,grh2010} or
the surface of molecular clouds \citep{wr2004}.

The Effelsberg telescope was used previously to map the Galactic emission
up to medium latitudes at 1.4~GHz with an angular resolution of
$9\farcm4$ \citep{ufr99,rei04}. Part of the data were combined with
polarization observations from the interferometric Canadian Galactic
Plane Survey (CGPS) \citep{lrr2010} at $1\arcmin$ angular
resolution. At 1.4~GHz, mostly local polarized emission is observed
\citep{gdm01,ulg03}. The Westerbork Synthesis Radio Telescope maps at
350~MHz \citep{wdj93,hkd03a,hkd03b} penetrate through an even shorter
distance into the magneto-ionic medium. Earlier 2.7~GHz Galactic plane
polarization surveys were made with the Effelsberg 100~m telescope
\citep{jfr87,drr99} with a $4\farcm3$ resolution tracing more distant
emission. The Galactic plane in the southern sky was surveyed by the
Parkes telescope at 2.4~GHz \citep{dhj97}, while sections were
observed with the Australian Telescope Compact Array at 1.4~GHz with
a high angular resolution \citep{gdm01,hgm06}. A review of polarization
surveys and their calibration was given by \citet{rei06}. To penetrate
even deeper into the magnetoionic medium, observations at higher
frequencies are required.

The Sino-German $\lambda$6~cm polarization survey covers the Galactic
plane from $10\degr\leq l\leq230\degr$ and $|b|\leq5\degr$ with an
angular resolution of $9\farcm5$. The detailed description of the
survey strategy was presented in \citet[][Paper~I]{shr07} for a
test region of $122\degr\leq l\leq129\degr$. The observations and
results for the anti-centre region of $129\degr\leq l \leq230\degr$
were published by \citet[][Paper~II]{grh2010}, and the results
for the region of $10\degr\leq l \leq60\degr$ were discussed in
some detail by \citet[][Paper~III]{srh2010}. In this paper, we present
the results for the remaining region of $60\degr\leq l\leq122\degr$,
where the line-of-sight is directed almost parallel to the local arm
at lower longitudes and almost across the outer arms at higher
longitudes. The large-scale Galactic magnetic fields in the arm and
inter-arm regions \citep{hq1994,hml2006} cause significant Faraday
rotation for polarized emission at 4.8~GHz at low Galactic
longitudes. For completeness, we include the maps of the test region
($122\degr\leq l\leq129\degr$) discussed in Paper~I. We briefly
describe the observations, data reduction, and the zero-level
restoration for polarized emission in Sect.~2.  In Sect.~3, we discuss
extended sources and polarized structures, including
polarized patches, \ion{H}{II} regions acting as Faraday
screens, as well as the fluctuation properties of diffuse polarized
emission of the entire survey region.  A summary is given in
Sect.~4. The survey data will be released at the ``MPIfR Survey
Sampler"\footnote{http://www.mpifr.de/survey.html} and the web-page of
the $\lambda$6~cm survey at NAOC\footnote{http://zmtt.bao.ac.cn/6cm/}.

\section{Observations and data reduction}

The Sino-German $\lambda$6~cm polarization survey of the Galactic
plane has been performed with the 25-m telescope of the Urumqi Observatory,
National Astronomical Observatories of the Chinese Academy of Sciences
at Nanshan station. The $\lambda$6~cm receiver is a copy of a receiver
used at the Effelsberg 100-m telescope and was installed at the Urumqi
telescope in August 2004. A detailed description of the receiving
system and the observation modes was presented in Paper~I.  In
brief, the system temperature $T_\mathrm{sys}$ is about 22~K. The
receiving system was tuned to a central frequency of 4800~MHz with a
bandwidth of $\Delta \nu=$ 600~MHz. Later a second set-up with a
central frequency of 4963~MHz with a narrow bandwidth of $\Delta \nu=$
295~MHz could be chosen to avoid interference by Indian geostationary
satellites when observing close to their positions.

\begin{figure}[htbp]
\includegraphics[angle=-90,width=0.48\textwidth]{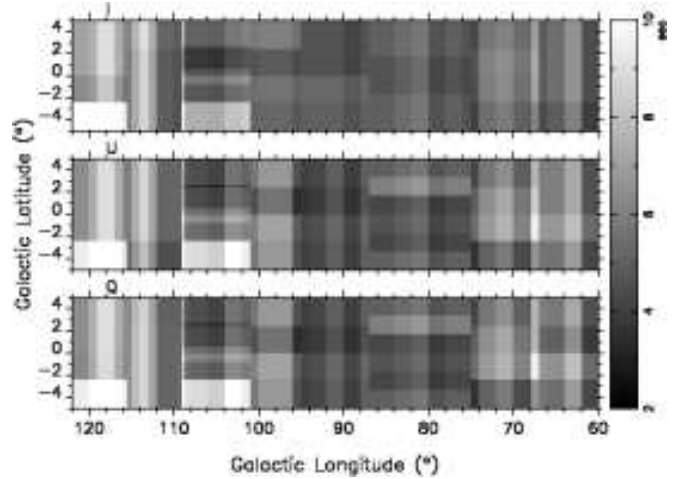}
\caption{The distributions of effective integration time for the
  $\lambda$6~cm survey of the Galactic plane in the region of $60\degr\leq l
  \leq122\degr$ for Stokes $I$, $U$ and $Q$ maps (from {\it top} to
  {\it bottom}).  }
\label{rms}
\end{figure}

The survey observations of the Galactic plane were started in autumn
2004 and completed in April 2009.  The Galactic plane was scanned in the 
direction of either Galactic longitude or Galactic latitude with a velocity
of $3\degr-4\degr$ per minute to ensure that the instrumental and weather
conditions remained stable enough during observations of a single field. 
See details in Paper~III. The separation between subsequent scans was
$3\arcmin$ providing full sampling for a HPBW of $9\farcm5$.  Each
individual survey field was scanned at least six times when using the
narrow band to achieve high sensitivity. 3C286 served as the main
calibrator with an assumed flux density of 7.5~Jy, and a 11.3\% linear
polarization at a polarization angle of $33\degr$. Both 3C138 and 3C48
served as secondary calibrators. We always observed one calibrator
before and after mapping a survey field. The conversion factor between
flux density and main beam brightness temperature was 0.164~K/Jy.

The observed data were processed in several steps. We first removed
visible interferences, then adjusted ground radiation and elevation-dependent 
atmospheric base-level distortions of each individual scan by applying 
a second-order polynomial fit, and removed scanning effects in the
$I$, $U$, and $Q$ maps by applying the ``unsharp masking" method of
\citet{sr79}. The baselines for scans observed in the Galactic
longitude direction were restored by the corresponding map measured
along the Galactic latitude direction.  These corrections were applied
to all $I$, $U$, and $Q$ maps. Remaining pointing errors were
calibrated by comparing point source positions from the survey with
those from the high-angular resolution NVSS catalog
\citep{ccg1998}. Finally, all maps were combined using the
PLAIT algorithm \citep{er88}. The instrumental polarization, mainly
the leakage of total power emission into the polarization channels,
was determined by observations of the unpolarized calibrators 3C295
and 3C147, and corrected using the ``REBEAM" method as explained
in Paper~I.

The effective integration time $t$ (i.e. scaled to the bandwidth of
600~MHz) for $I$, $U$, and $Q$ maps is shown in Fig.~\ref{rms} for the
present survey region. For observations with a bandwidth of
$\Delta\nu=$ 600~MHz and three coverages, the expected rms-noise
$\sigma=T_\mathrm{sys}/\sqrt{\Delta\nu t}$ is 0.9~mK $T_\mathrm{B}$
for $I$, and $\sigma_{U/Q}$=0.6~mK $T_\mathrm{ B}$ for both $U$ and $Q$.
The measured rms-values for emission-free regions in general agree
with this estimate.

\begin{figure*}[!htbp]
\begin{center}
\includegraphics[angle=-90,width=0.98\textwidth]{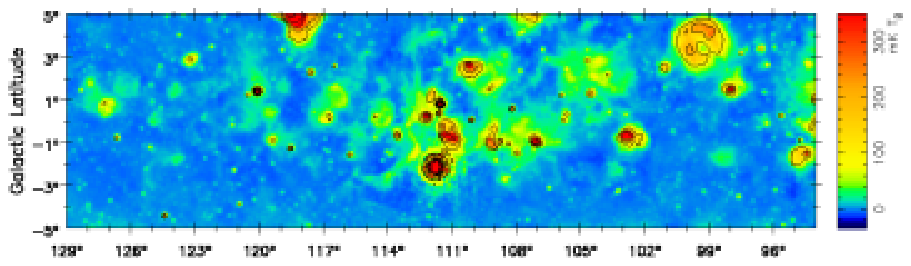}
\includegraphics[angle=-90,width=0.98\textwidth]{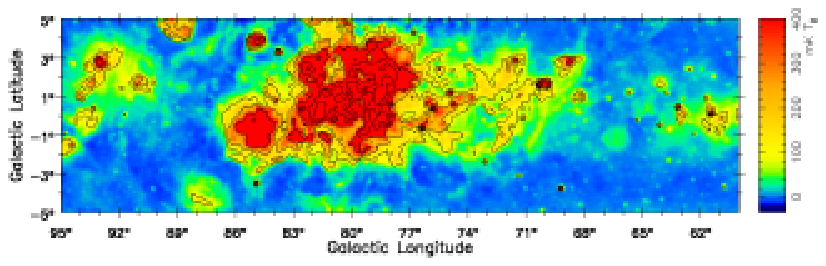}
\caption{Total intensity maps for the survey region of $60\degr \leq l
  \leq 129\degr$.  Overlaid total intensity contours are shown at
  $2^n\times100$~mK~$T_\mathrm{B}$ with $n=0,\,1,\,2,\,\ldots$.}
\label{6cm.i}
%
\includegraphics[width=0.23\textwidth,angle=-90]{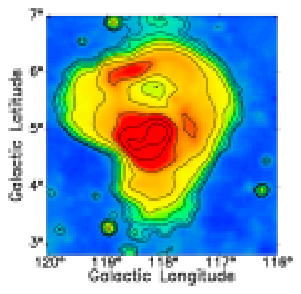}
\includegraphics[width=0.23\textwidth,angle=-90]{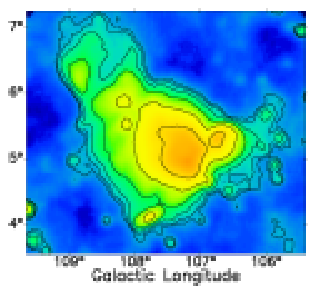}
\includegraphics[width=0.23\textwidth,angle=-90]{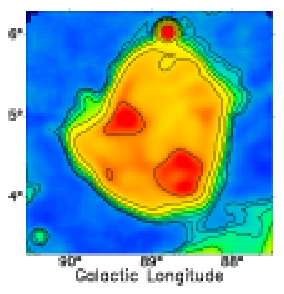}
\includegraphics[width=0.23\textwidth,angle=-90]{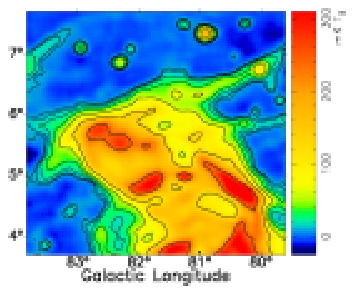}
\caption{Four extended sources partly located outside the
  high-latitude boundary of the survey maps.  From {\it left} to {\it
 right}, we show the \ion{H}{II} regions W1 and CTB107 and the SNRs
  HB21 and W63. Overlaid total intensity contours are at
  $6.0+2^n\times3.6$~mK~$T_\mathrm{B}$ with $n=0,\,1,\,2,\,\ldots$.
}
\label{6cm.sou}
\end{center}
\end{figure*}

\begin{figure*}[htbp]
\begin{center}
\includegraphics[angle=-90,width=0.95\textwidth]{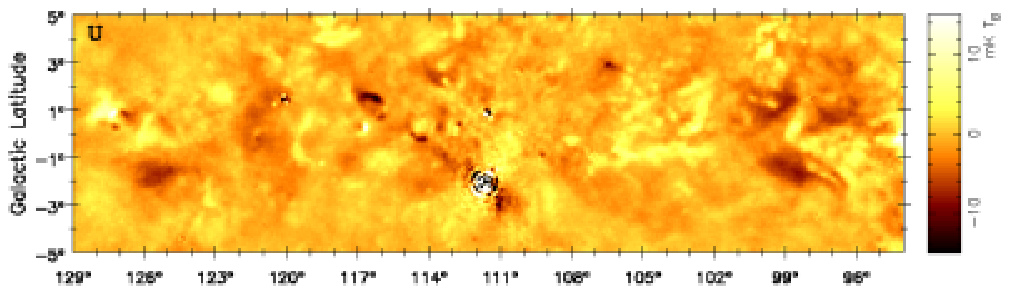}
\includegraphics[angle=-90,width=0.95\textwidth]{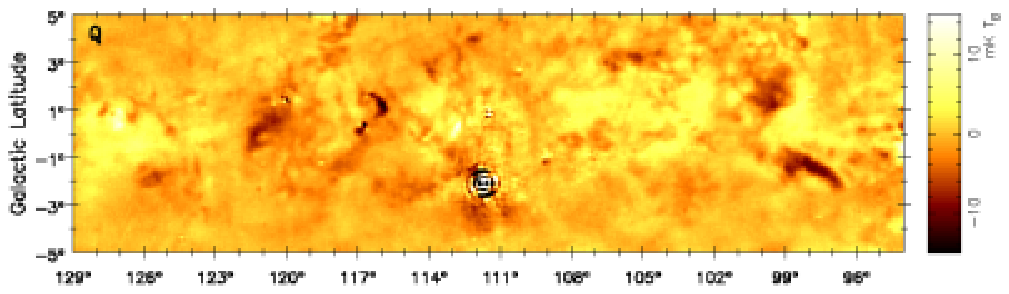}
\includegraphics[angle=-90,width=0.95\textwidth]{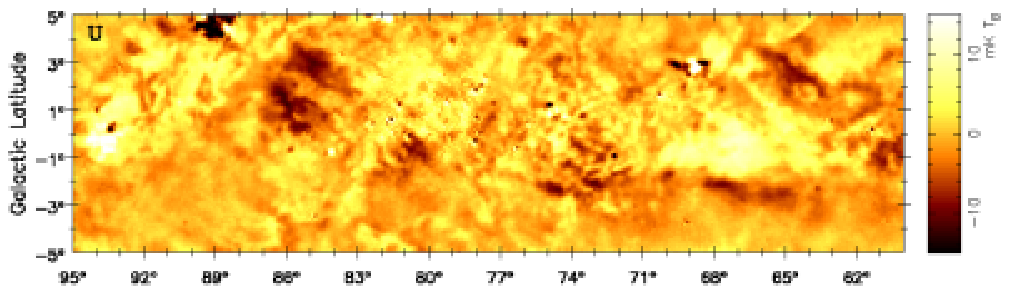}
\includegraphics[angle=-90,width=0.95\textwidth]{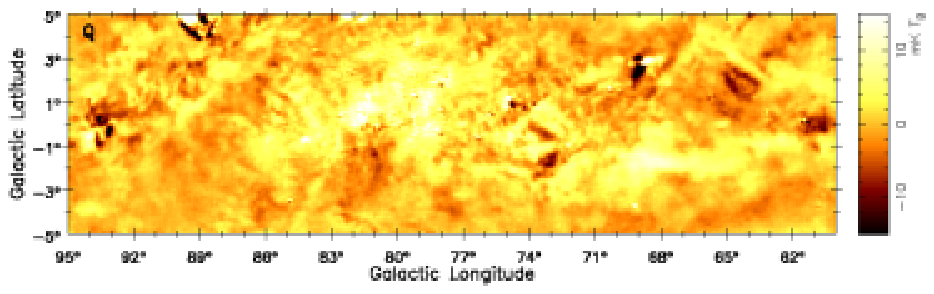}
\caption{Observed Stokes $U$ and $Q$ maps for the
region of $60\degr \leq l \leq 129\degr$.
}
\label{6cm.uq}
\end{center}
\end{figure*}

\section{Results}

We first present the total intensity maps, followed by the
polarization maps. We briefly discuss interesting objects visible in
the $\lambda$6~cm survey, i.e. \ion{H}{II} regions and Faraday
screens, and finally investigate the fluctuation properties of the
magnetized interstellar medium. Many known discrete objects,
such as supernova remnants and HII regions, are clearly visible in the
present $\lambda$6~cm survey region of $60\degr \leq l \leq
122\degr$. Supernova remnants are distinctive because of their polarized synchrotron
emission. Known large SNRs ($>1\degr$) will be discussed by Gao et al.
(2011, Paper~V), and small diameter SNRs by Sun et al. (2011, in
prep.).

\subsection{Total intensity maps}

The total intensity maps for the region of $60\degr\leq l
\leq129\degr$, $|b|\leq5\degr$ are shown in Fig.~\ref{6cm.i}. Each map
covers a region of $35\degr \times 10\degr$ with an overlap of
$1\degr$ in longitude. The total intensity maps in this area show
prominent structures of star-forming regions, \ion{H}{II} regions, and
SNRs. The overall radio structures in the total intensity map is quite
similar to what is seen in the Effelsberg $\lambda$21\ cm
\citep{kr1980,rrf90,rrf1997} and $\lambda$11\ cm survey maps
\citep{rfrr90,frr90}. The most outstanding region is the strong Cygnus
complex region located at about $77\degr<l<87\degr$ and
$-2\degr<b<4\degr$, where indistinguishable radio emission from
\ion{H}{II} regions and other objects accumulates in the tangential
direction of the local arm \citep[see the distribution of \ion{H}{II}
  regions in][]{hhs2009}. It is almost entirely thermal emission of
objects located at distances between 1~kpc and 4~kpc
\citep{whl1991}. Diffuse emission from the inner disk ($l<65\degr$)
originates almost entirely from the Sagittarius arm
\citep{hhs2009}. Extended arcs, e.g. the one at $l=114\fdg0$,
$b=-2\fdg5$, or filaments, e.g. the one a $l=87\fdg8$, $b=3\fdg5$, are
clearly visible.
Antenna sidelobes show up in the area of Cas~A at $l=111\fdg7$,
$b=-2\fdg1$, where data were difficult to clean.
Four extended sources located partly beyond the high-latitude boundary
of this survey section are the two \ion{H}{II} regions, W1
(G118.2+5.0) and CTB107 (G107.4+5.2), and the two SNRs, HB21 and
W63. We observed them separately 
and show the resulting maps in Fig.~\ref{6cm.sou}.

\subsection{Polarization maps}

The observed $U$ and $Q$ maps are presented in Fig.~\ref{6cm.uq}.  We
detected various polarized structures of different scales in the
survey regions, some stretching out from the Galactic disk.  The
$\lambda$6~cm maps miss emission components of scales larger than
about $10\degr$, because we set the two ends of the latitude subscans
to zero.  However, absolute base-levels for $U$ and $Q$ are important
to calculate correct polarization angles and intensities. Missing
large-scale components lead to a misinterpretation of polarization
data \citep{rei06,lrr2010}.  For the
$\lambda$6\ cm survey, the ongoing C-Band All- Sky Survey
(CBASS)\footnote{http://www.astro.caltech.edu/cbass} might provide the
missing large-scale emission. CBASS aims for a sensitivity of about
0.1 mK, which is very similar to that of the Urumqi $\lambda$6~cm
polarization survey when smoothed to the CBASS angular resolution of
$44'$.  However, CBASS data are not available now.

The polarized emission at both the WMAP K-band (22.8~GHz) and the
$\lambda$6~cm band originate from synchrotron emission (see
discussions in Paper III). The best we can do at present is to
extrapolate the large-scale $U$ and $Q$ components at $\lambda$6~cm
from the WMAP K-band (22.8~GHz) polarization data
\citep{phk2007,hwh2009}, which have an absolute zero-level. The
spectral index of the polarization intensity or the synchrotron
emission in the plane varies from $\beta=-3.1$ to $\beta=-2.7$ from
$l=10\degr$ to $l=60\degr$ (Paper~III), and becomes $\beta=-2.9$ from
$l=129\degr$ to $l=230\degr$ (Paper~II). Here $\beta$ is the
brightness temperature spectral index defined as $T_B = \nu^{\beta}$,
where $\nu$ is the observing frequency.  A clear trend of steepening
of the spectra in the region of $105\degr<l<120\degr$ is indicated in
the spectral index map between 408~MHz and 1420~MHz obtained by
\citet{rr88a,rr88b}. We therefore model the spectral index along the
Galactic plane as: (1) $\beta=-2.7$ for the region of
$60\degr<l<105\degr$; (2) $\beta=-2.9$ for the region of
$120\degr<l<129\degr$; and (3) linear interpolation between $\beta=-2.7$
and $\beta=-2.9$ for the region of $105\degr<l<120\degr$.  The $U$ and
$Q$ maps at 22.8~GHz were scaled to $\lambda$6~cm (4.8~GHz) according
to the spectral index $\beta$.

\begin{figure}[btp]
\centering
\includegraphics[width=0.38\textwidth,angle=-90]{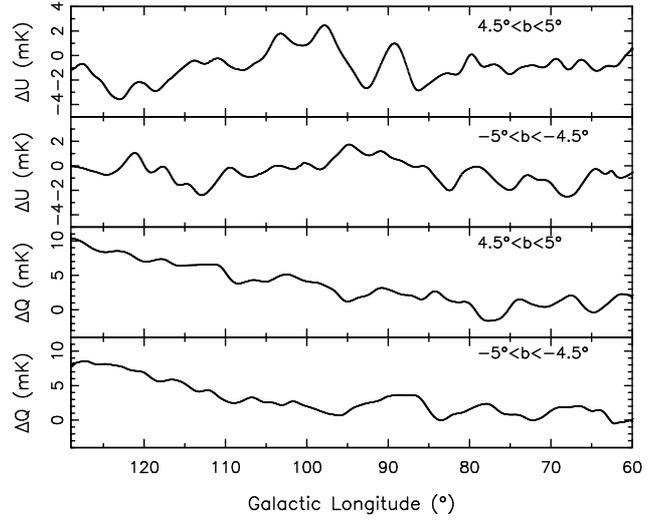}
\caption{Difference between the $U$ and $Q$ maps at
    $\lambda$6\ cm extrapolated from the WMAP five-year K-band 
    at 22.8~GHz and the observed $U$ and $Q$ maps at 
    $\lambda$6\ cm for the high latitude areas. }
\label{restore}
\end{figure}

As discussed in Paper~III, the polarization horizon at $\lambda$6\ cm
in directions near $b=0\degr$ for the Galactic longitude range of
$10\degr<l<60\degr$ is about 4~kpc on average, much less than the
polarization horizon at the 22.8~GHz K-band and larger than the size of
the Galaxy. At larger longitudes of the survey region in this paper,
the polarization horizon around $l=100\degr$ is quite comparable to
that at the 22.8~GHz K-band, even near $b=0\degr$.  However, for
longitudes around $l=70\degr$, the horizons are different.  To ensure
that the emission at $\lambda$6~cm and the 22.8~GHz K-band originates
from the same volume, we calculated the differences between the $U$
and $Q$ maps extrapolated from the 22.8~GHz K-band and the observed
data only in the high-latitude regions of $4\fdg5<|b|<5\degr$
(Fig.~\ref{restore}).

\begin{figure}[tb]
\centering
\resizebox{0.45\textwidth}{!}{\includegraphics[angle=-90]{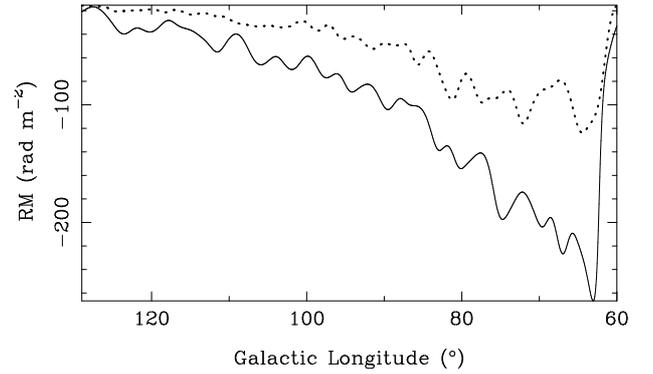}}
\caption{Simulated average RM profiles for the area 
  $-5\degr<b<-4\fdg5$ (solid line) and $4\fdg5<b<5\degr$ (dotted line)
  along the Galactic longitude range $60\degr<l<129\degr$.}
\label{rm}
\end{figure}

However, the RM in this region could be large in general so that a
simple extrapolation of the $U$ and $Q$ data from the 22.8~GHz K-band
to $\lambda$6~cm (4.8~GHz) may not be correct. A correction for RM
should be applied.  The RMs were estimated from simulations of $U$ and
$Q$ maps at the $\lambda$6~cm band and the 22.8~GHz K-band by using
the \textsc{hammurabi} code of \citet{wjr+09} and the 3D-emission
models by \citet{srwe08}, which properly reproduce the observed total
intensity, polarization, and RM surveys. From the simulated
polarization angle maps, the RM map were obtained. The RM profiles in
the regions of $-5\degr<b<-4\fdg5$ and $4\fdg5\degr<b<5\degr$ are
shown in Fig.~\ref{rm}.

\begin{figure*}[htbp]
\begin{center}
\includegraphics[angle=-90,width=0.95\textwidth]{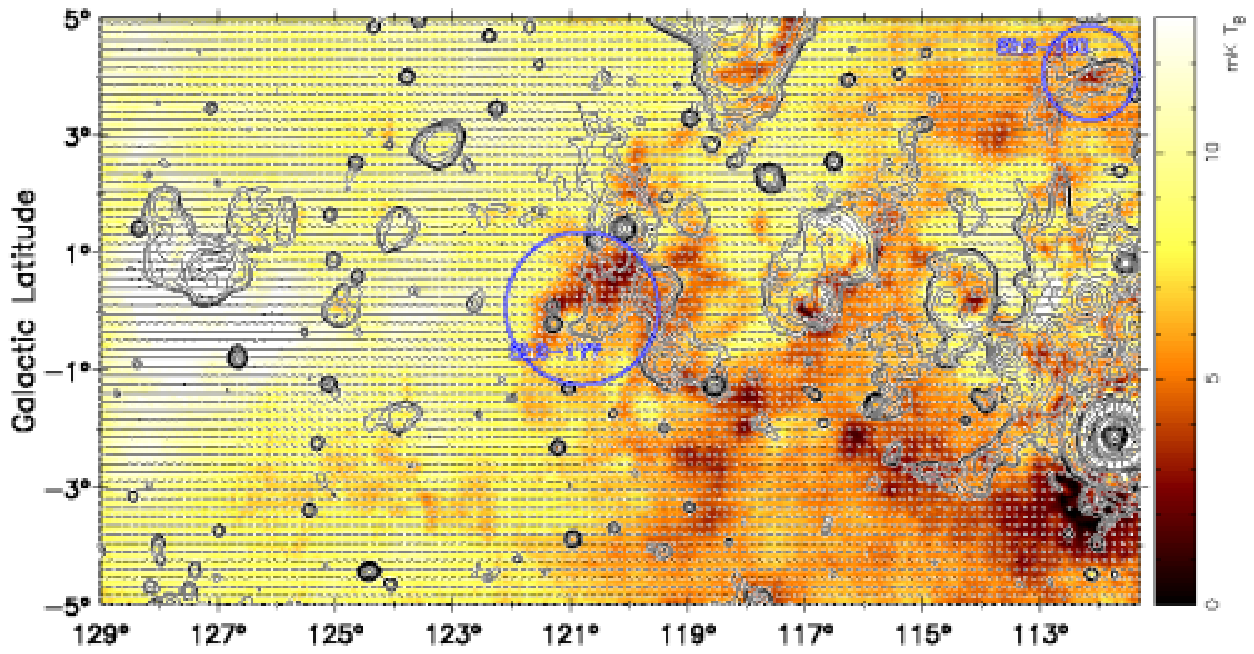}
\includegraphics[angle=-90,width=0.95\textwidth]{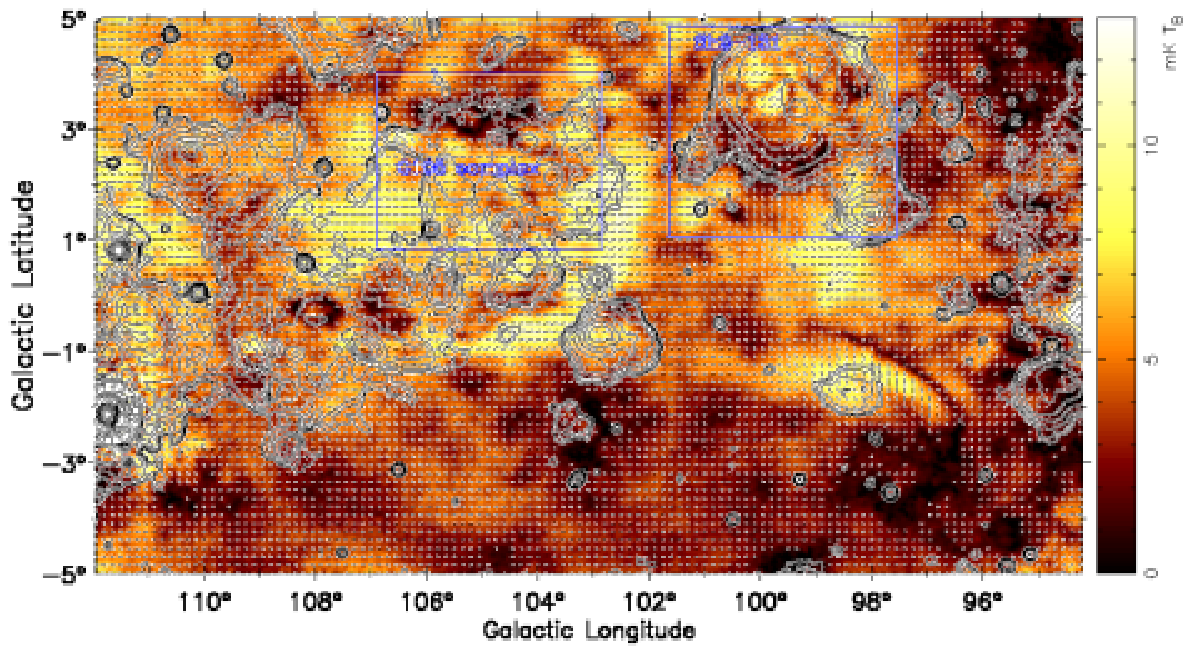}
\caption{Zero-level restored $PI$ maps for the region of $94\degr \leq
  l \leq 129\degr$. The superimposed bars are separated by 9$\arcmin$ and
  are shown in the B-vector direction with lengths proportional to $PI$
  and a lower intensity cutoff of 2.5~mK $T_\mathrm{B}$ ($5\times
  \sigma_{PI}$). Contours show total intensities running from 6.0~mK
  $T_\mathrm{B}$, in steps of 3.6$\times 2^n$ mK $T_\mathrm{B}$ (n =
  0,1,2,$\ldots$).  Boxes mark strong depolarization along the
    periphery of the \ion{H}{II} regions, Sh2-131 and the G105 complex. 
    Circles mark the depolarization \ion{H}{II} regions, Sh2-177 and
    Sh2-160.}
\label{6cm.add.l}
\end{center}
\end{figure*}

\begin{figure*}[htbp]
\begin{center}
\includegraphics[angle=-90,width=0.95\textwidth]{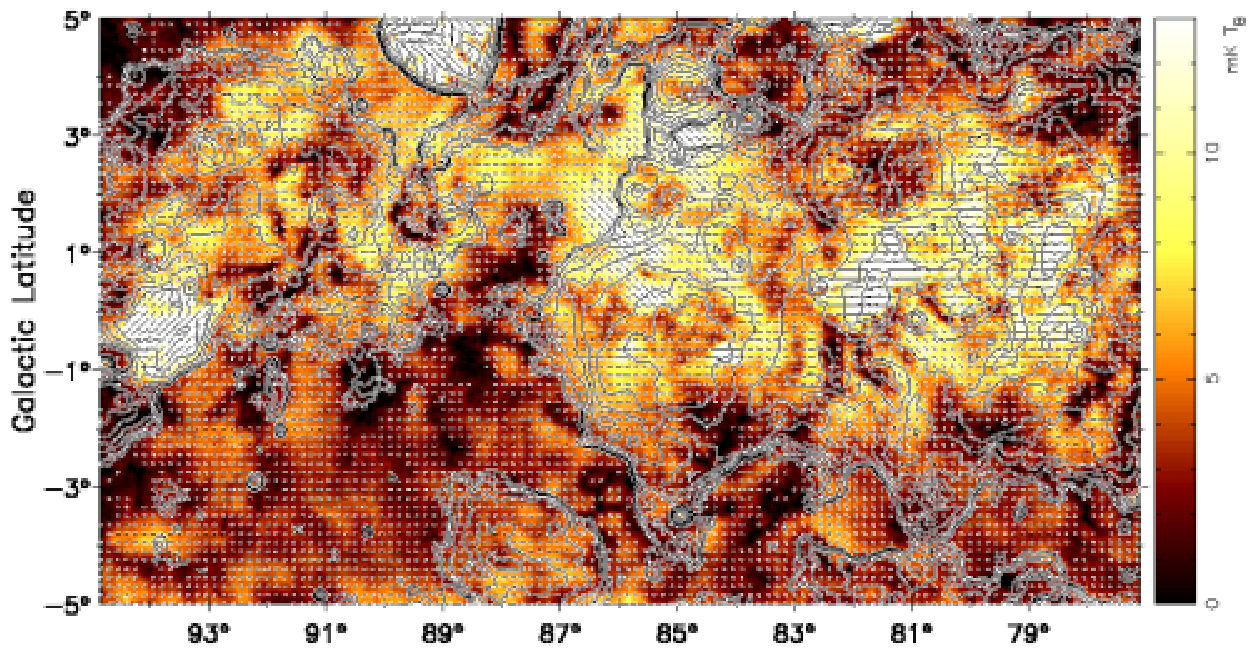}
\includegraphics[angle=-90,width=0.95\textwidth]{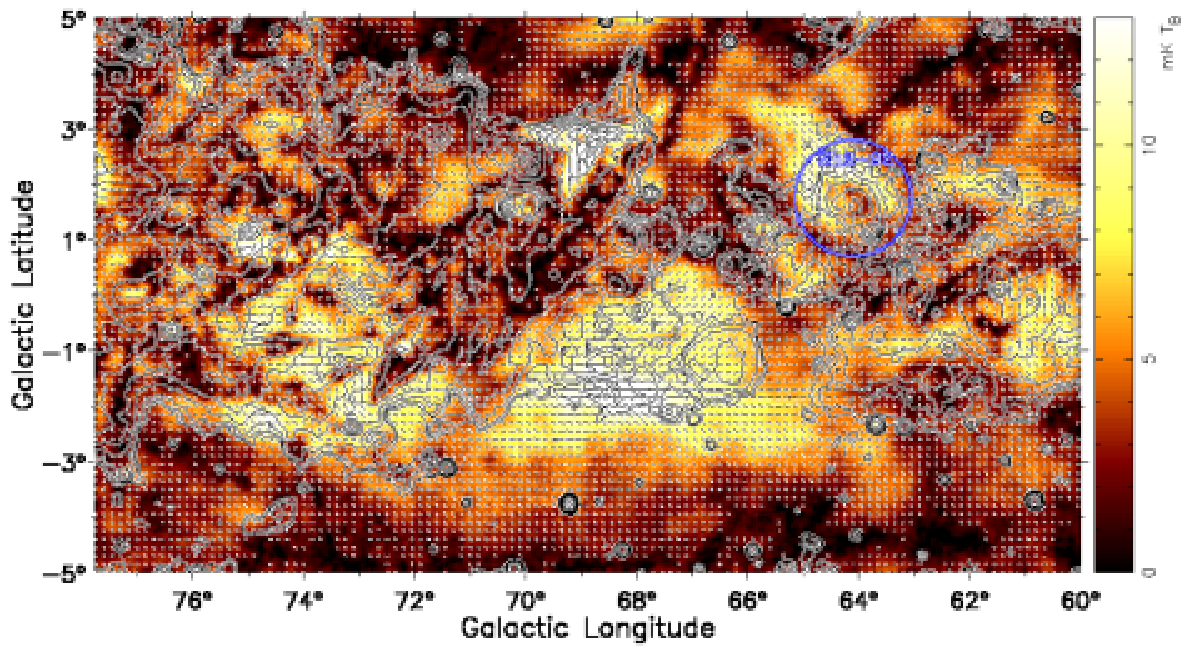}
\caption{The same as Fig.~\ref{6cm.add.l} but for the region of
  $60\degr \leq l \leq 95\degr$. The depolarization around the
  \ion{H}{II} region Sh2-92 is indicated by a circle.}
\label{6cm.add.r}
\end{center}
\end{figure*}

With the extrapolated $\lambda$6\ cm $U$ and $Q$ maps from the WMAP
22.8~GHz K-band in the two high latitude regions and the simulated RM
map as described above, we performed the $U$ and $Q$ restoration of
the $\lambda$6~cm maps. Both the extrapolated maps and the observed
$\lambda6$~cm (4.8~GHz) maps were convolved to the same resolution of
$2\degr$. Their difference at the high latitude regions and the linear
interpolations for lower latitudes are added to the observed $U$ and
$Q$ maps for the restoration. Two strong extended
polarized sources SNRs HB~21 (G89.0+4.7) and W~63 (G82.2+5.3) were
excluded before smoothing the data to $2\degr$ angular resolution. For
the maps in the region of $122\degr<l<129\degr$, which were
presented in Paper~I, the polarization data were restored for the
absolute levels for $U$ and $Q$, where a spectral index of
$\beta=-2.9$ and no RM corrections were applied. With the new
restorations, the results for the two Faraday screens G124.9$+$0.1 and
G125.6$-$1.8 discussed in Paper~I were only marginally changed,
because of the relatively small RM corrections towards
$122\degr<l<129\degr$ (see Fig.~\ref{rm}).

A restoration of the large-scale polarized emission does not change
the polarization structures in $U$ and $Q$ much, but could
significantly change polarized intensity ($PI$) and polarization
  angle ($PA$, see Paper~I, II and III). We note that the restoration
has some uncertainties. The spectral index $\beta$ for the polarized
emission is uncertain by $\Delta\beta=0.4$ for the inner region of
$60\degr<l<105\degr$ as discussed in Paper~III, which introduces a
maximal error of 1.8~mK~$T_{\rm B}$ for zero-level restoration. For
the outer region of $120\degr<l<129\degr$, the spectral index
uncertainty drops to $\Delta\beta=0.1$ (Paper~II), but missing
emission in $Q$ is large (see Fig.~\ref{restore}) causing a maximal
error of 2~mK~$T_{\rm B}$. In both cases, the uncertainties in the
restoration process are less than 4$\times\sigma_{PI}$.  Therefore, our
restoration is a good approximation. 

\begin{figure*}[!bt]
\begin{tabular}{ccc}
\includegraphics[angle=-90,width=0.30\textwidth]{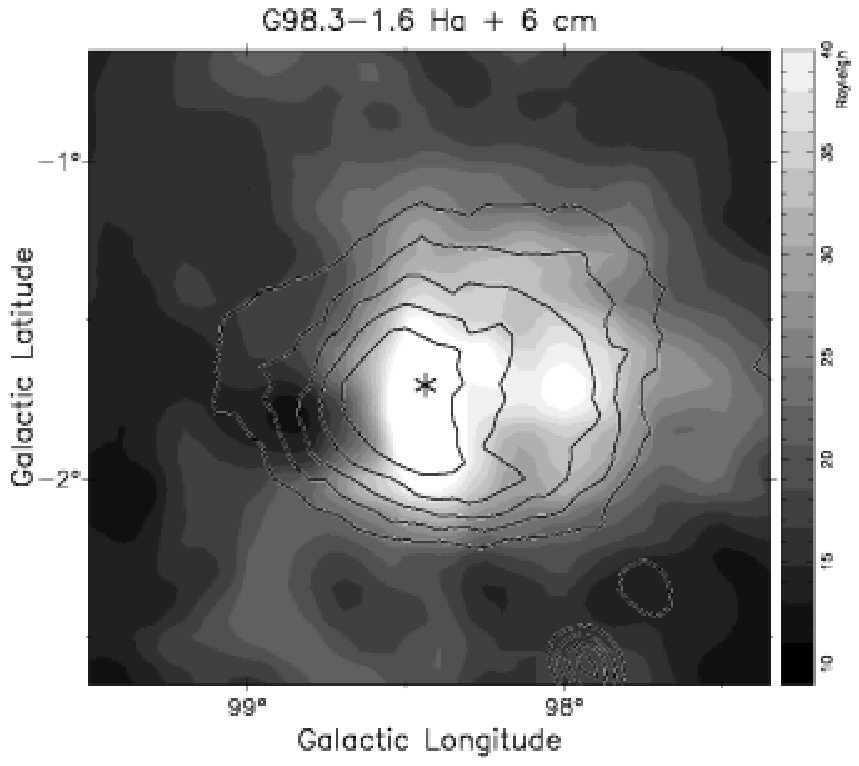} &
\includegraphics[angle=-90,width=0.29\textwidth]{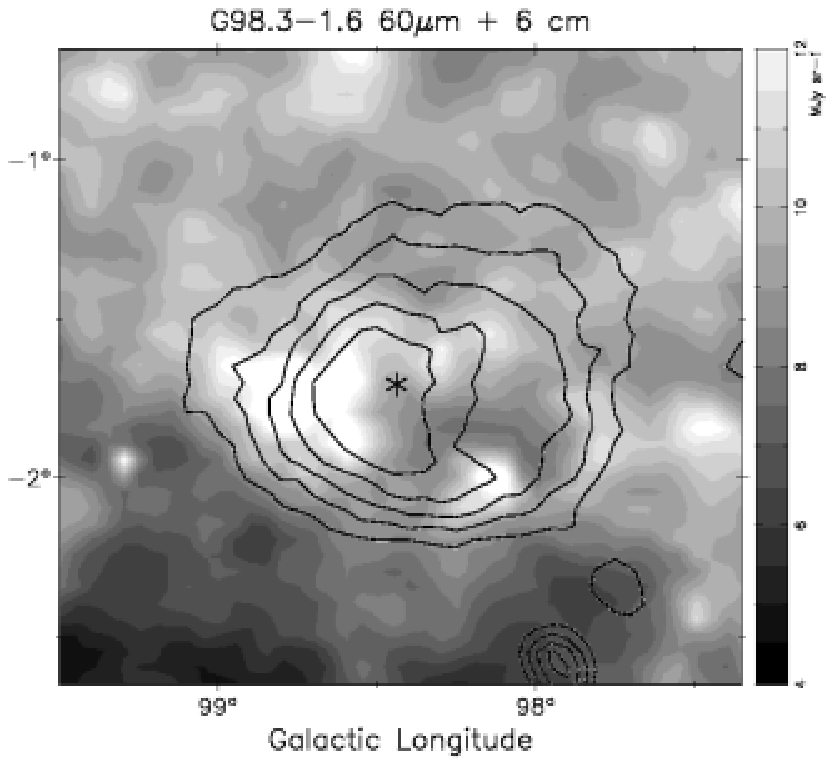} &
\includegraphics[angle=-90,width=0.33\textwidth]{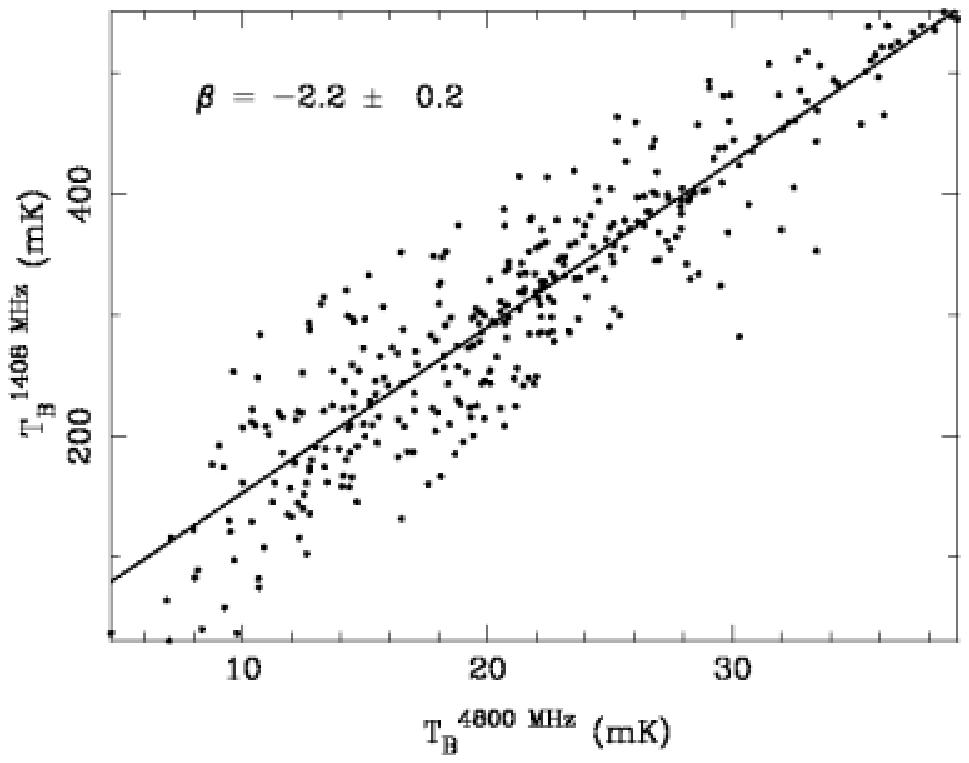} \\
\includegraphics[angle=-90,width=0.30\textwidth]{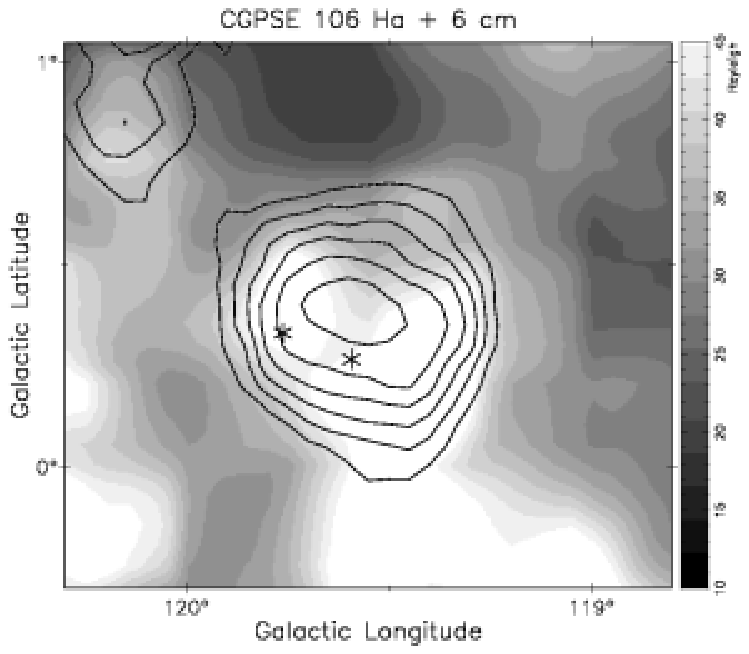} &
\includegraphics[angle=-90,width=0.29\textwidth]{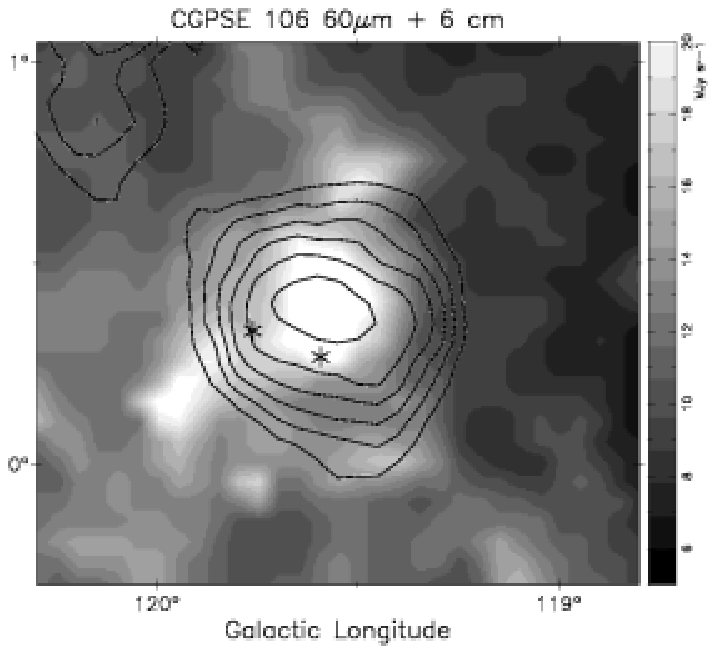} &
\includegraphics[angle=-90,width=0.33\textwidth]{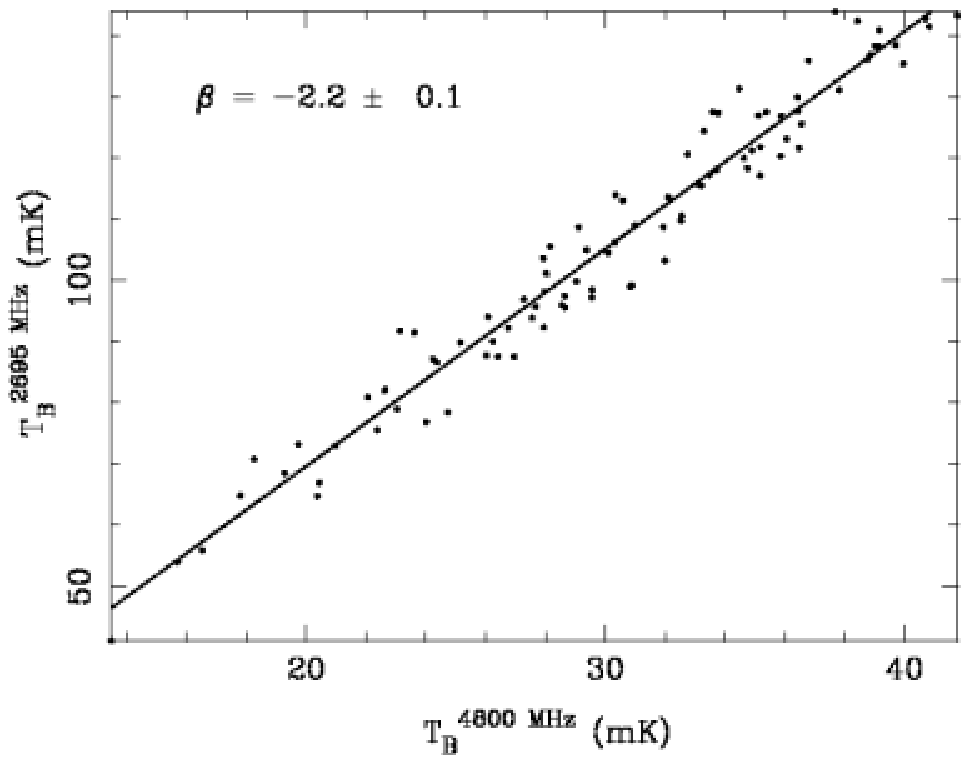}
\end{tabular}
\caption{Two newly identified extended \ion{H}{II} regions:
  G98.3$-$1.6 ({\it upper} panels) and CGPSE 106 ({\it lower}
  panels). Total intensity contour maps at $\lambda$6\ cm were
  overlaid onto corresponding H$\alpha$ images \citep[{\it left}
    panels,][]{f2003} and infrared (60~$\mu$m) images \citep[{\it
      middle} panels,][] {mml2005}.  Contours of total intensity at
  $\lambda$6\ cm start from 5~mK $T_\mathrm{B}$ and run in steps of
  5~mK $T_\mathrm{B}$. The B-stars in the fields were marked by
  stars. TT-plots of the $\lambda$6~cm data and and lower frequency
  Effelsberg survey data ({\it right} panels) are used to obtain
  spectral indices.}
\label{newHII}
\end{figure*}
 
We present $PI$ maps in Figs.~\ref{6cm.add.l} and \ref{6cm.add.r}
derived from the restored large-scale $U$ and $Q$ maps. The diffuse
polarization intensity becomes stronger towards larger Galactic
longitude, indicating that depolarization is more significant at smaller
longitude, especially for the region of $l<97\degr$. The highly
polarized `Fan region' also contributes for $l>120\degr$. Some
prominent discrete polarized structures appear in the
original Stokes $U$ and $Q$ maps, e.g. the spurs at $l=98\fdg2$,
$b=-1\fdg6$ and $l=64\fdg8$, $b=2\fdg4$ and complicated structures
towards the Cygnus complex region. In the region of $l>97\degr$,
after adding the large-scale polarized emission, some polarized
features appear as a minimum region in the $PI$ map with significant
variation in the polarization angles (e.g. the region around
$l=120\fdg7$, $b=0\fdg4$).

In the region of $l<97\degr$ a few large polarized patches, and
depolarized regions with ``depolarized canals'' appear in the restored
polarization map. The origin of canal structures in our $\lambda$6~cm
maps has been discussed in Paper~III. Depolarization has by several
causes \citep{burn66,sbs98}. Polarized synchrotron emission from the
turbulent and tangled magnetic fields in different layers along the
line-of-sight are superimposed and naturally result in depolarization. The
amount depends on the intrinsic properties of the magnetized
interstellar medium. The Cygnus region is complicated and turbulent.
Large-scale magnetic fields being parallel to the line-of-sight
cause significant Faraday rotation. Polarized emission from
different depths undergo a different amount of Faraday rotation and
destructively reduce the observed level of polarization. In addition, the
average polarized emission from slightly different directions
within the beam will reduce the degree of polarization. The thermal
gas is dense with a small filling factor, so that any polarized
emission from areas beyond the arm suffers depolarization. Therefore,
the detected polarization emission is probably produced in a region
nearer than the Cygnus-X region at 1~kpc~\citep{whl1991}.

\subsection{Two new \ion{H}{II} regions}
 
\ion{H}{II} regions are ionized regions with strong H$\alpha$
emission and a high thermal electron density. They have a flat
spectrum and related emission features from dust in the infrared
bands.  Some \ion{H}{II} regions act as Faraday screens (see Paper~I,
II and III) hosting a strong regular magnetic field along the
line-of-sight, as we discuss below.
 
In this survey region, we discovered two new extended \ion{H}{II}
regions (see Fig.~\ref{newHII}). The first one is G98.3$-$1.6 with a
size of $90\arcmin\times66\arcmin$.  Strong H$\alpha$ emission
\citep{f2003} is associated (see Fig.~\ref{newHII}), but no H$\alpha$
source in this area was catalogued previously. One B-type star
\citep[$>$1.5 kpc,][]{r2005} is located near the $\lambda$6~cm and
H$\alpha$ emission peaks. The flux density we measured at
$\lambda$6~cm is $4.4\pm0.3$~Jy. The flux contributions from
large-scale diffuse emission and unresolved extragalactic sources have
been excluded. The TT-plot spectral index, $\beta$ (defined as
  $T_B = \nu^{\beta}$) derived from the $\lambda$6~cm data and the
Effelsberg $\lambda$21\ cm survey data is $\beta_{6/21}=-2.2\pm0.2$
(see Fig.~\ref{newHII}).

The second source is CGPSE 106 (G119.6+0.4), which was found as an
extended source in the CGPS 1.4~GHz survey \citep{kmp2007}. It has a
size of $44\arcmin$. H$\alpha$ and infrared emission \citep[see
  Fig.~\ref{newHII}][]{f2003,mml2005} was detected in this region. Two
B-type stars are located near the centre, one of which (LSI +62 91)
has a distance of 3.38~kpc \citep{kw1989}. We subtracted extragalactic
sources within the CGPSE 106 region from the Effelsberg $\lambda$11~cm
and $\lambda$21~cm maps and got integrated flux densities of $S_{\rm
  11cm}=2.4\pm0.5$~Jy and $S_{\rm 21cm}= 2.6\pm0.6$~Jy. From the
$\lambda$6~cm map, we obtained $S_{\rm 6cm}= 2.3\pm0.3$~Jy. The TT-plot
spectral indices for the diffuse emission are $\beta_{6/11}=-2.2 \pm
0.1$ and $\beta_{6/21}=-2.1 \pm 0.5$.  We conclude that both
G98.3$-$1.6 and CGPSE 106 are \ion{H}{II} regions.

\begin{figure}[!hbt]
\begin{center}
\includegraphics[angle=-90,width=0.42\textwidth]{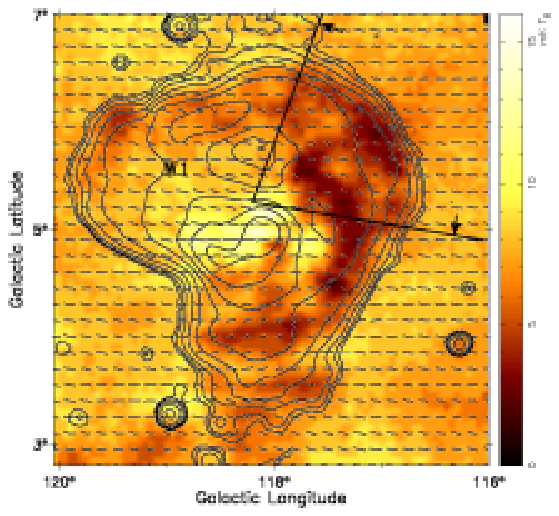}
\includegraphics[angle=-90,width=0.42\textwidth]{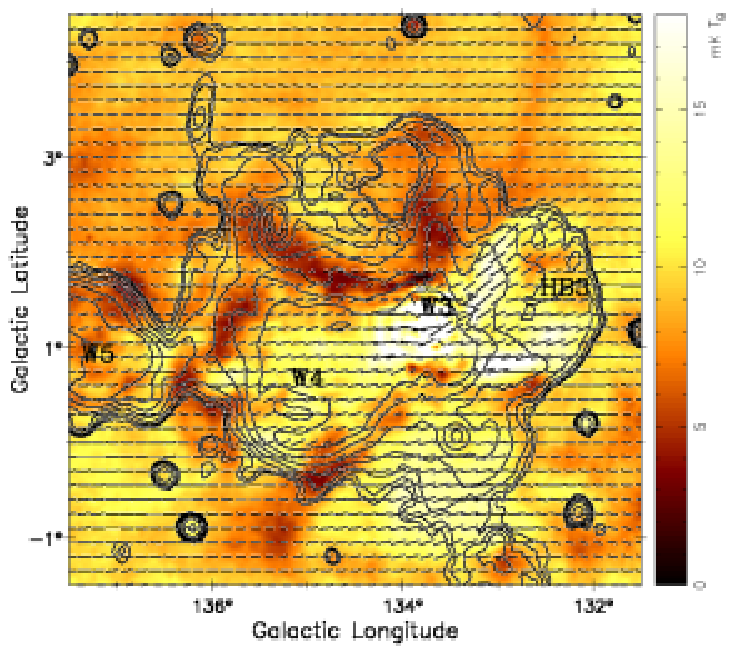}
\includegraphics[angle=-90,width=0.42\textwidth]{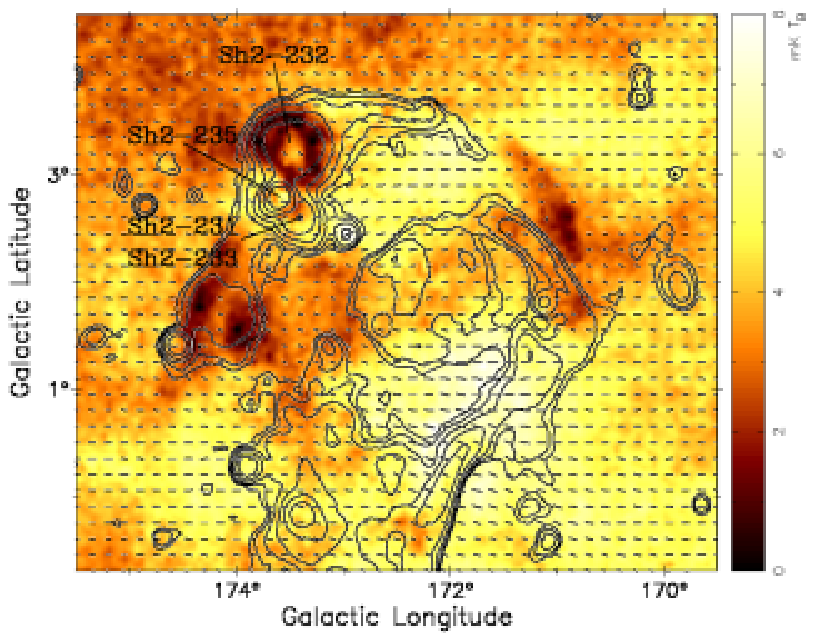}
\caption{As Fig.~\ref{6cm.add.r} but for W1 ({\it top}),
    W3/W4/W5 ({\it middle}) and the G173 complex ({\it bottom})
    published in Paper~II showing depolarization along the
    periphery of \ion{H}{II} complexes. Depolarization of the
    indicated sector (position angles between $-16\degr$ and
    $-97\degr$) in the outer periphery of W1 is modeled in Sect.~3.4.
  Total intensity contours of the W1, W3/W4/W5 and G173 complex start
  at 6.0, 6.0, and 3.0~mK~$T_\mathrm{B}$, respectively, and increase in
  steps of $2^n\times3.6$~mK~$T_\mathrm{B}$, with $n=0,\,1,\,2,\,\ldots$.}
\label{w1}
\end{center}
\end{figure}

\subsection{Depolarization along the periphery of \ion{H}{II} region complexes}

Strong depolarization is visible along the periphery of a number of
\ion{H}{II} region complexes. The apparent widths of these narrow
depolarization arcs are $15\arcmin$ to $25\arcmin$. Quite prominent
examples are the southern and western periphery of Sh2-131 near
$l=99\fdg3$, $b=3\fdg7$, or the northern periphery of the G105
\ion{H}{II} complex regions near $l=105\fdg0$, $b=2\fdg6$ as seen in
the restored polarization maps in Fig.~\ref{6cm.add.l}. The \ion{H}{II} region  W1 (see
Fig.~\ref{w1}) also shows depolarization along its western
periphery. Two more clear examples are seen in the outer Galaxy, which
were published in Paper~II. These are the W3/W4/W5 complex at
$l=135\fdg0$, $b=0\fdg8$ and the G173 \ion{H}{II} region complex at
$l=172\fdg8$, $b=1\fdg8$.  We replotted the corresponding polarization 
intensity maps in Fig.~\ref{w1}.

Sh2-131 (see Fig.~\ref{6cm.add.l}) is an extended \ion{H}{II} region
(2\degr$\times$2\degr) excited by the star cluster Trumpler~37, which
has a distance of 0.86~kpc~\citep{bfs82}.  The depolarized arc at
$l=100\fdg0$, $b=2\fdg2$ appears along the outer boundary of the lower
part. The polarization intensity of the arc decreases to 2~mK
$T_\mathrm{B}$, while the general polarization level is about 6~mK
$T_\mathrm{B}$.

The G105 complex (see Fig.~\ref{6cm.add.l}) contains the \ion{H}{II}
region Sh2-134 at 0.9~kpc~\citep{bfs82} and SNR G106.3$+$2.7 at
0.8~kpc ~\citep{kup2001}.  The nearly totally depolarized region is
located along the northern (upper) boundary of the complex.

The \ion{H}{II} region W1 (see Fig.~\ref{w1}) has a complex shell structure with a size of
$3\degr\times3\degr$. Its distance is 850~pc \citep{m68} and it is
excited by the star cluster Be~59, containing one O7 star and several
later types stars. Depolarized patches are detected inside W1.  The
largest depolarized area in the western part has two arcs, with a
length of about $1\fdg5$ each (see Fig.~\ref{w1}). The outer one runs
along the periphery. The inner arc is just outside an emission ridge
(see Figs.~\ref{6cm.sou} and \ref{w1}). The polarization intensity
decreases to about 4 mK $T_\mathrm{B}$ compared to the polarized
background of about 8 mK $T_\mathrm{B}$.

The \ion{H}{II} regions W3, W4, and W5, together with the SNR HB3,
form a prominent emission complex in the Perseus arm at a distance of
about 2~kpc. W4 is ionized by the star cluster IC 1805. It is
apparently connected with W5 in the east (i.e. $l=137\fdg6$,
$b=1\fdg1$) and W3 in the west (i.e. $l=132\fdg7$,
$b=1\fdg3$). Clearly visible in Fig.~\ref{w1} is a continuous
depolarized feature along the northern, eastern, and southern periphery
of W4 at a level of about 4~mK $T_\mathrm{B}$ compared to the on-average 
high polarization intensity of 10~mK $T_\mathrm{B}$ in the
surroundings or even higher close to W3.

The G173 complex consists of several \ion{H}{II} regions (Sh2-232,
Sh2-235, Sh2-231, and Sh2-233) at a distance of 1.8~kpc~\citep{hcl1996}
and was discussed in Paper~II. We outline the associated
depolarized structures in Fig.~\ref{w1}, where the strongest
depolarization is caused by Sh2-232 at $l=173\fdg5$, $b=3\fdg2$. One
depolarized arc is located west of the radio ridge at l=$171\fdg0$,
b=$2\fdg7$. Another depolarized region appears within the inner
periphery of the eastern ridge at l=$174\fdg3$, b=$1\fdg9$.

All these depolarized structures close to \ion{H}{II} regions are
believed to be caused by Faraday screens, which rotate polarized
background emission, and cause depolarization compared to their
surroundings when adding to the unrotated polarized foreground
emission. \citet{grh2010} discussed in some detail the Faraday screen
properties for the \ion{H}{II} region W5. Here we model the
depolarization effect along the periphery of the \ion{H}{II} region W1
as an example.

\begin{figure}
\begin{center}
\includegraphics[angle=-90,width=0.46\textwidth]{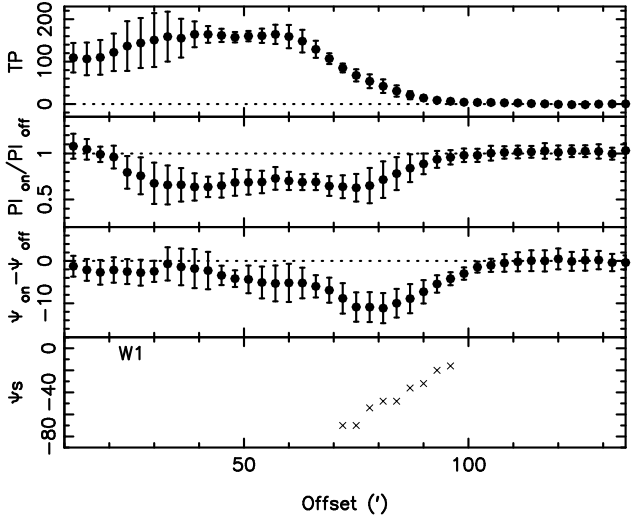}
\caption{The averaged radial distributions of the total power (TP,
  {\it top}), $PI$ ratio ({\it upper middle}), and the $PA$ ({\it lower
    middle}) difference in the sector region of W1 defined in
  Fig.~\ref{w1}. The calculated rotation angle $\psi_{s}$ caused by
  the Faraday screen is also shown ({\it bottom}). }
\label{w1_par}
\end{center}
\end{figure}

We use the Faraday screen model that was described in detail in
Paper~I. We assume that the polarized background emission is smooth on
scales larger than the object and that polarization angles for
background and foreground emission are the same. This simplification
is required for single frequency observations, but, as already
discussed in Paper~II, seems to be quite valid at least for outer
Galaxy directions.  The polarization differences between the Faraday
screens ``on" and ``off" position can be described as (see Paper~I)
\begin{equation}\label{pipa_onoff}
\setlength\arraycolsep{0.2pt}
\left\{
\begin{array}{rcl}
\displaystyle{\frac{PI_{\rm on}}{PI_{\rm off}}}&=&
\sqrt{f^2(1-c)^2+c^2+2fc(1-c)\cos2\psi_{\rm s}},\\[2mm]
\psi_{\rm on}-\psi_{\rm off}&=&\displaystyle{\frac{1}{2}\tan^{-1}
\frac{f(1-c)\sin2\psi_{\rm s}}{c+f(1-c)\cos2\psi_{\rm s}}},
\end{array}
\right.
\end{equation}
Where $f$ is the depolarization factor by the Faraday screen ranging
from 0 (total) to 1 (no depolarization), $c=PI_{\rm fg}/(PI_{\rm
  fg}+PI_{\rm bg})$ is the fraction of foreground polarization, and
$\psi_{s}=RM\cdot\lambda^{2}$ is the angle rotated by the Faraday
screen.

For a sector of W1 in the western periphery, between a position
angle of $-16\degr$ and $-97\degr$ as indicated in Fig.~\ref{w1}, we
obtained the averaged radial distribution (Fig.~\ref{w1_par}) of $PI$
and $PA$. By model-fitting, we calculated the averaged depolarization
factor to be about $f=0.8$ and $c=0.73$. We measured rotation angles
$\psi_s$ between $-20\degr$ and $-70\degr$ (see Fig.~\ref{w1_par}).
At $\lambda$6~cm, the maximum rotation angle of $-70\degr$ requires
$RM=-339$~rad m$^{-2}$ for this direction of the \ion{H}{II} region
shell.  For the 850~pc distance of W1, a shell thickness of
$20\arcmin$ and a shell radius of $95\arcmin$, we calculated a maximum
path length $L=29$~pc.
We simply considered a uniform electron density $n_e$ in the shell.
For free-free emission, the electron density $n_e$ can be derived from the
$\lambda$6~cm brightness temperature $T_{s}=\tau T_{e}$ assuming an
electron temperature $T_e=8000$~K, which is typical of Galactic \ion{H}{II} regions. 
The optical depth $\tau$ is calculated following \citet{rw2000} as 
\begin{equation}\label{tau}
\tau=8.235\times10^{-2}\big(\frac{T_{e}}{\rm K}\big)^{-1.35}
\big(\frac{\nu}{\rm GHz}\big)^{-2.1}\big(\frac{EM}{\rm pc~cm^{-6}}\big),
\end{equation}
where $EM=n_{e}^{2}L$ is the emission measure depending on the path
length $L$ and the electron density $n_e$. We calculated $\tau$ from
the $\lambda$6~cm total intensity of 90~mK\ $T_\mathrm{B}$ and
obtained a maximum $EM\sim 682$~pc cm$^{-6}$ for W1.  The electron
density is then on average $n_{e}\sim 4.8$~cm$^{-3}$.
Combining the path length $L$, electron density $n_e$, and modeled RM,
we calculated a magnetic field strength along the line-of-sight in the
shell of $B_{||}\sim 3.0~\mu$G\index{}.

\subsection{Depolarization by small diameter \ion{H}{II} regions}

\ion{H}{II} regions are characterized by an enhanced thermal electron
density, and may also harbor some regular magnetic fields. They may
act as Faraday screens as discussed above and also in Papers~I
and II.  Here, we focus on the depolarization properties of four
\ion{H}{II} regions in the present survey region, which are not well
resolved by our $9.5'$ beam.  The objects are Sh2-92 at
$l=64\fdg1$, $b=1\fdg7$, BFS 13 at $l=108\fdg1$, $b=-0\fdg4$, Sh2-160
at $l=112\fdg2$, $b=4\fdg0$, and Sh2-177 at $l=121\fdg1$, $b=0\fdg4$.

\medskip
\noindent{\bf Depolarization by Sh2-92:} A broad spur-like feature
showing strong polarized emission runs from about $l=65\fdg5$,
$b=3\fdg5$ to $l=63\fdg5$, $b=0\fdg5$ (Fig.~\ref{6cm.add.r}). The spur
is probably a discrete feature (see discussions in next
subsection). Here we discuss the \ion{H}{II} region Sh2-92 with a
distance of 4.4~kpc \citep{fh81_2}, as indicated by the $\lambda$6\ cm
total power contours, which matches the depolarized area at the root
of this spur. The polarization intensity decreases to about 5 mK
$T_\mathrm{B}$, compared to the polarized emission from the spur of
about 12 mK $T_\mathrm{B}$. We note that the PA values of the spur
vary smoothly over the depolarization region, which suggests that the
polarized spur is an emission structure that is nearer than the \ion{H}{II}
region Sh2-92. If the spur were further away than Sh2-92, it would
have a physical size exceeding 240~pc. Since the emission from the
spur is an additional emission component for our Faraday screen model,
data at just one frequency are insufficient to apply the model. At
the periphery of the polarized ridge at $\lambda$6\ cm, a polarized
emission ridge at $\lambda$11\ cm runs from about $l=66\fdg0,
b=4\fdg0$ to $l=63\fdg5, b=2\fdg0$ in the Effelsberg $\lambda$11~cm
polarization survey map \citep{drr99}. Most of the $\lambda$6~cm spur
emission seems to be depolarized at $\lambda$11~cm. Towards Sh2-92, no
polarized emission is seen at $\lambda$11~cm. The depolarization
towards Sh2-92 observed at $\lambda$6~cm may therefore be caused by the
beam depolarization of polarized background emission, but almost
pure Faraday rotation also seems to be possible, leading to a reduced
polarization intensity from the spur by superposition.

\begin{figure}[hbt]
\centering
\includegraphics[angle=-90,width=0.42\textwidth]{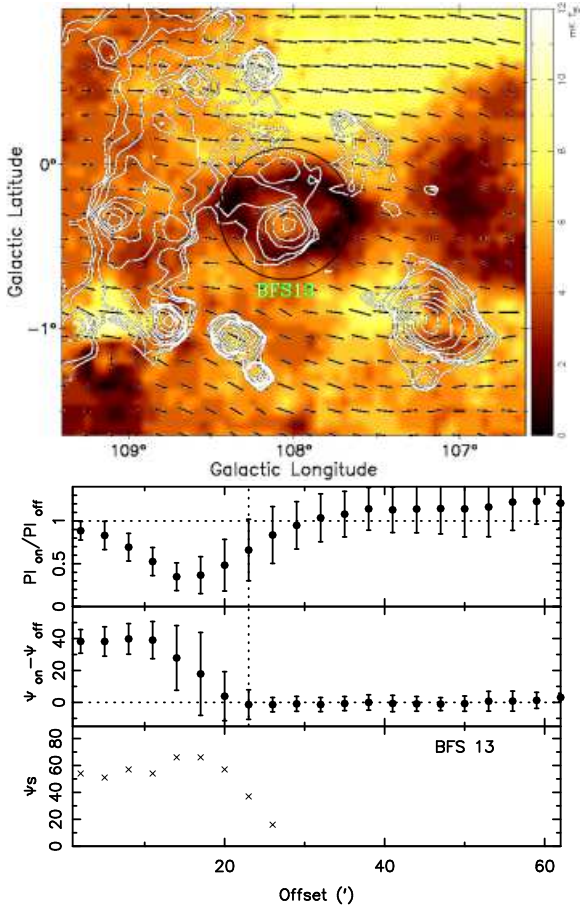}
\includegraphics[angle=-90,width=0.4\textwidth]{108FS.line.ps}
\caption{{\it Upper panel:} Thermal emission at 60~$\mu$m
  \citep{mml2005} shown by contours overlaid onto the image of
  polarization intensity of the area around BFS~13. {\it Lower panel:}
  $PI$ ratio and $PA$ difference for BFS~13 compared to its
  surroundings. The modeled rotated angles ($\psi_s$) of BFS~13
  increase to the maximum of $\sim 66\degr$ in the outer region.}
\label{108FS.plot}
\end{figure}

\medskip
\noindent{\bf Depolarization by BFS~13:} We detected depolarization by
the \ion{H}{II} region BFS~13 at $l=108\fdg1$, $b=-0\fdg4$
(Fig.~\ref{108FS.plot}). The distance of BFS~13 was estimated to be
about 1.4~kpc from the velocity of associated CO emission
\citep{bfs82}. The apparent diameter of the depolarized region is
about 48$\arcmin$, which corresponds to a linear size of 20~pc.  We
plotted the average radial distribution of polarization intensity $PI$
and polarization angle $PA$ in Fig.~\ref{108FS.plot}.  The
depolarization reaches a maximum at the periphery of BFS~13.  The
average polarization angles decrease from 30$\degr$ to 0$\degr$ in the
outer region.  Using the Faraday screen model, we calculated $f=0.8$,
$c=0.3$, and the rotation angle $\psi_{s}$ by the Faraday screen to be
$\sim 66\degr$, corresponding to a RM of about 320~rad m$^{-2}$. From
the peak brightness temperature of BFS~13 of about 28~mK
$T_\mathrm{B}$, we calculated an EM of 213~pc~cm$^{-6}$ according to
Eq.~(\ref{tau}), and an average electron density within BFS~13 of
about 3.3~cm$^{-3}$. The regular magnetic field along the
line-of-sight within BFS~13 is then $B_{||} \ge 6.0~\mu$G.

\medskip
\noindent{\bf Depolarization by Sh2-160:} Depolarization is seen
towards the \ion{H}{II} region Sh2-160 at $l=112\fdg2$, $b=4\fdg0$
(see Fig.~\ref{6cm.add.l}). This \ion{H}{II} region has a distance of
about 0.9~kpc \citep{bfs82}. The diameter of the depolarized region is
about 16$\arcmin$, corresponding to about 4.2~pc. The polarized
intensity decreases to 3~mk~$T_\mathrm{B}$ compared to
7~mk~$T_\mathrm{B}$ in its surroundings. The Faraday screen model
results are $f=1.0$, $c=0.4$, and $\psi_{s}=-67\degr$, which
corresponds to a RM of about $-325$~rad m$^{-2}$. For this almost
unresolved \ion{H}{II} region, from the peak brightness temperature of
Sh2-160 of about 32~mK $T_\mathrm{B}$, we calculated an EM of
243~pc~cm$^{-6}$ according to Eq.~(\ref{tau}), and get the 
electron density within Sh2-160 of about 7.6~cm$^{-3}$.  From the RM,
path length, and electron density, we estimated the regular magnetic
field strength along the line-of-sight as $B_{||}\sim 12~\mu$G.

\begin{figure}[hbt]
\centering
\includegraphics[angle=-90,width=0.42\textwidth]{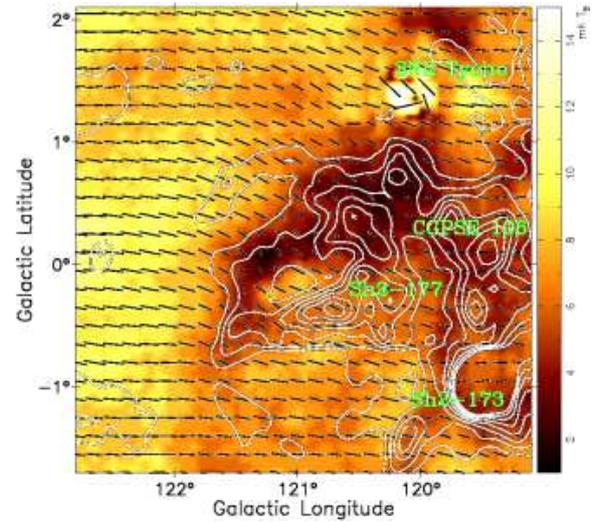}
\caption{H$\alpha$ emission \citep{f2003} of Sh2-177 is shown by contours
  starting at 8~Rayleigh and running in steps of 5~Rayleigh overlaid
  onto the image of the polarization intensity at $\lambda$6~cm.}
\label{FS120}
\end{figure}

\medskip
\noindent{\bf Depolarization by Sh2-177:} A large depolarized region
($2\degr \times \sim1\degr$) is detected in the north-east
(upper-left) extension of the \ion{H}{II} region Sh2-177 (see
Fig.~\ref{FS120}), which has a distance of 2.5~kpc~\citep{bfs82}. The
shape of the depolarized region agrees very well with the extended
H$\alpha$ emission region \citep{f2003} from Sh2-177, which indicates
the possible association. $PI$ decreases to 3~mK $T_\mathrm{B}$ over
an area of 30$\arcmin$. The enhanced H$\alpha$ emission is about
10~Rayleigh there. Using the Faraday screen model to fit the observed
deviations of $PA$ and $PI$ from its surroundings, we obtained
$f=1.0$, $c=0.5$, a maximum rotation by the Faraday screen of
$\psi_s=-70\degr$ corresponding to a RM of about $-339$~rad m$^{2}$.
For this region, we estimated the electron density from the associated
$\rm H\alpha$ intensity, $I_{\rm H\alpha}$. We first calculated the
emission measure \citep{f2003}
\begin{equation}\label{em}
EM=2.75T_4^{0.9}I_{\rm H\alpha}\exp\left[2.44E(B-V)\right],
\end{equation}
where $E(B-V)$ is the magnitude of reddening, which is about 1.1 in
this region \citep{sfd1998}. We got a value of $EM =
328$~pc~cm$^{-6}$.  The depolarized area has a size of 30$\arcmin$,
i.e. 22~pc, so that an average electron density of $n_e\sim
3.8$~cm$^{-3}$ results. The magnetic field strength $B_{||}$ is then
about $5.0~\mu$G.

\subsection{Polarized patches}

A variety of patchy polarized emission has been detected in this
region of the Galactic plane. Most of these patches do not have a
corresponding total intensity emission. The bright patches of
polarized emission have an intensity between 10~mK and
12~mK~$T_\mathrm{B}$ (see Fig.~\ref{6cm.add.l} and
Fig.~\ref{6cm.add.r}). The sizes of these patches range from about
30$\arcmin$ up to several degrees. Here we list three prominent bright
polarized patches.

\begin{enumerate}

\item A polarized spur was detected around $l=64\fdg5, b=2\fdg0$ with
  a length of $4\degr$ and an inclination of $45\degr$ to the Galactic
  plane. The polarization angles almost uniformly follow this feature
  indicating a very uniform magnetic field.  As already noted above,
  the spur is largely depolarized at $\lambda$11~cm.

\item A large polarized plume at $l=68\fdg5, b=-1\fdg5$ has a size of
  $7\degr \times 3\degr$ with polarization angles mostly orientated
  parallel to the Galactic plane. Similar but narrower polarization
  structures are seen in the Effelsberg 2.7~GHz survey~\citep{drr99}
  and the high-resolution 1.4~GHz survey by \citet{lrr2010}.

\item The polarized patch around $l=86\fdg0, b=1\fdg5$ has an
  irregular shape, a size of $4\degr \times 4\degr$, and almost
  uniform polarization angles of $+45\degr$ in its central region. It
  is composed of two parts with a depression in the centre region with
  a size of $2\degr \times2\degr$.
\end{enumerate}

These prominent polarized structures all have a large angular size of
a few degrees in the sky. They should be discrete physical structures
with sizes of several tens of pc. We noticed that the nearby prominent
\ion{H}{II} region W80 is located within the polarized patch around
$l=86\fdg0, b=1\fdg5$, which has a distance of 850~pc \citep{wbb1983},
but has little effect on the polarized structure. The polarized patch 
is most likely produced by synchrotron emission in a region nearer
than 850~pc. We assume that all these patches have a size smaller than 50~pc,
the polarized spur $l=64\fdg5, b=2\fdg0$ with an angular size of
$4\degr$ should have a distance of $<$715~pc; the polarized plume at
$l=68\fdg5, b=-1\fdg5$ of size $7\degr$ should have a distance of
$<$410~pc; the polarized patch around $l=86\fdg0, b=1\fdg5$ of
$4\degr$ should lie within 715~pc.  Therefore, they probably originate
from synchrotron emission in a nearby emission region with a well-ordered 
magnetic field of typical strength, because no excessive
synchrotron emission is observed in total intensity.

\subsection{Structure function analysis for polarization data}

Polarization data for all survey regions of $10\degr<l<230\degr$ are
now available (Paper~II, III, and this paper), we are able to
discuss them in context.  We clearly see an increase in polarization
fluctuations towards lower Galactic longitudes.  The $\lambda$6~cm
survey with a beam of $9\farcm5$ should in principle reveals all
fluctuations on scales between $30\arcmin$ and $3\degr$,
which are related to the turbulent properties of the magnetized
interstellar medium. \citet{srh2010} analyzed
polarization fluctuations using the structure functions of $PI$, $Q$, and $U$
maps for the survey region of $10\degr<l<60\degr$.  Here, we analyse
the polarization data of all survey regions of $10\degr<l<230\degr$
to investigate how the fluctuations change with Galactic longitude.

The second-order structure function for polarization intensity is
  calculated with
\begin{equation}
F_{PI}(\Delta
\theta)=\frac{\sum_{i=1}^N[PI(\theta_i)-PI(\theta_i+\Delta
    \theta)]^{2}}{N},
\label{structfun}
\end{equation}
where $\Delta \theta$ is the angular separation between the two pixels,
and $N$ is the number of pixel pairs with the same
$\Delta\theta$. Structure functions for $U$ and $Q$ were calculated in
a similar way, with underlying noise subtracted \citep{sts11,
    hgm04}.  We fit the structure function with a power law
\begin{equation}
F = A \; (\Delta \theta)^n
\end{equation}
for angular scales between $30\arcmin$ and $3\degr$. Here $A$ represents
for the fluctuation power of polarized structures, and $n$ is the 
power-law index for the fluctuation power distribution on different
scales.

\begin{figure}[hbt]
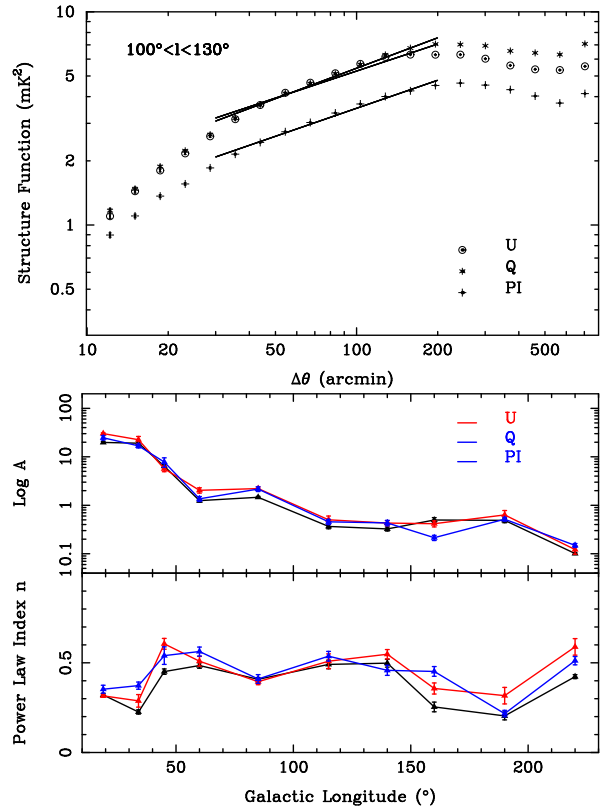

\begin{center}
\includegraphics[angle=-90,width=0.43\textwidth]{100-130.rms.ps} \\
\includegraphics[angle=-90,width=0.43\textwidth]{sf.color.new.ps}
\caption{{\it Upper panel:} Structure functions from the original
  $\lambda$6~cm $U$, $Q$, and $PI$ data for the region of
  $100\degr<l<130\degr$.  {\it Lower panel:} Amplitude and the
    power-law-index of the $U$, $Q$, and $PI$ structure
  functions change with Galactic longitude. 
}
\label{sf}
\end{center}
\end{figure}

We divided the observed $U$, $Q$, and $PI$ images into a number of
small sections according to the spiral tangents in the inner Galaxy
and the distribution of strong objects in the outer arm
\citep{hhs2009}. We calculated the structure functions from the
$U$, $Q$, and $PI$ maps for each region. One example for the region
$100\degr<l<130\degr$ is shown in Fig.~\ref{sf}. Obvious coherent
structures from SNRs were blanked in the maps before any
structure function calculations were made. The structure functions
always have a power law index $n$ around 0.5, which does not
change significantly with Galactic longitude. However, the
fluctuation power $A$ decreases with Galactic longitude
(Fig.~\ref{sf}).  More fluctuation power is detected at smaller
longitudes for the arm regions in the inner Galaxy, as expected.  In
the anti-centre region of the Galaxy, polarized structures are less
numerous, as reflected by the smaller fluctuation power $A$. For
the regions $10\degr\leq l\leq40\degr$ and $160\degr\leq l\leq
200\degr$, the power law indices are smaller than those in other
regions, which means relatively more polarized structures on smaller
angular scales. In the inner Galaxy, the enhanced fluctuations on
small scales may be caused by higher turbulence in the interstellar
medium in the arms \citep{hgb+06,hbg+08}.  In the anti-centre region
of $160\degr\leq l\leq 200\degr$, the large-scale magnetic fields
\citep{hq1994,hml2006} are almost perpendicular to the
line-of-sight. The enhanced fluctuations there are indicative of the dominant
random magnetic fields in the region.

\section{Summary}

We have surveyed the Galactic plane for continuum and polarized
emission at $\lambda$6~cm using the Urumqi 25-m telescope. In this
paper, we have presented and analysed the data for the region of
$60\degr\leq l \leq129\degr$, $|b|\leq5\degr$, where we see emission
in the outer Galaxy mainly from the Perseus arm and in the inner
Galaxy mainly from the Sagittarius arm.  We have restored the missing
large-scale $U$ and $Q$ in the survey maps by extrapolating the WMAP
five-year K-band data of~\citet{hwh2009}.

This $\lambda$6~cm survey provides new data for the study of discrete
objects. We identified two new HII regions. From a few large resolved
\ion{H}{II} region complexes, we detected coherent depolarization along
their periphery. We applied a Faraday screen model to these large
\ion{H}{II} regions and also some unresolved \ion{H}{II} regions
causing depolarization, and obtained their line-of-sight magnetic
field strength. Most of them are of the order of several $\mu$G.

We detected a few large polarized patches with an angular size of a
few degrees, which are probably discrete features within 1~kpc.

We have studied the fluctuation properties in the $U$, $Q$, and $PI$
maps in the survey region of $10\degr\leq l \leq230\degr$ by the
structure function analysis. Although the power law indices for
the structure functions are always close to about 0.5, the fluctuation
power is much lower for large longitudes or the anti-centre
region. More enhanced polarized structures on small scales give
much more fluctuation power towards the inner Galaxy.

The $\lambda$6~cm polarization survey is not only valuable for tracing
the Galactic magnetic field, but also provides important information
on magnetic fields within Galactic objects.

\begin{acknowledgements}
We thank Mr. XuYang Gao for helpful discussions and careful reading of
the manuscript and Dr. Weibin Shi for conducting some
observations. The Sino-German $\lambda$6\ cm polarization survey was
carried out with a receiver system constructed at MPIfR with financial
support by the MPG and the NAOC. We thank Mr. Maozheng Chen and Jun Ma
for operation support and maintenance. The survey team was supported
by the National Natural Science foundation of China (10773016,
10821061, and 10833003) and the National Key Basic Research Science
Foundation of China (2007CB815403) and the Partner group of the MPIfR
at NAOC in the frame of the exchange program between MPG and CAS for
many bilateral visits. XHS acknowledges financial support by the MPG
and by Prof. Michael Kramer during his stay at MPIfR.
\end{acknowledgements}

\bibliographystyle{aa}

\bibliography{gp}

\end{document}